\documentclass[aps,prl,showpacs,superscriptaddress,twocolumn]{revtex4}
\usepackage{amsfonts}
\usepackage{txfonts}
\usepackage{amssymb}
\usepackage{amsbsy} 
\usepackage{epsfig}

\def\be{\begin{equation}} \def\ee{\end{equation}}
\def\bea{\begin{eqnarray}} \def\eea{\end{eqnarray}}

\def\nn{\nonumber}

\def\bsigma{{\boldsymbol \sigma}}

\def\bQ{{\bf Q}}
\def\bk{{\bf k}}

\def\bx{{\bf x}}

\def\bp{{\bf p}}

\def\bB{{\bf B}}

\def\bK{{\bf K}}
\def\be{{\bf e}}

\def\la{\langle}
\def\ra{\rangle}

\begin{document}

\title{ Helical Spin Order from Topological Dirac and Weyl Semimetals}

\author{Xiao-Qi Sun}
\affiliation{ Institute for Advanced Study, Tsinghua University,
Beijing,  China, 100084}

\author{Shou-Cheng Zhang}
\affiliation{ Department of Physics, Stanford University, Stanford,
CA 94025}
\affiliation{ Institute for
Advanced Study, Tsinghua University, Beijing,  China, 100084}

\author{Zhong Wang}
\altaffiliation{Corresponding author.  wangzhongemail@tsinghua.edu.cn} \affiliation{ Institute for
Advanced Study, Tsinghua University, Beijing,  China, 100084}

\affiliation{Collaborative Innovation Center of Quantum Matter, Beijing 100871, China }

\date{ \today}

\begin{abstract}

We study dynamical mass generation and the resultant helical spin
orders in topological Dirac and Weyl  semimetals, including the
edge states of quantum spin Hall insulators, the surface states of
weak topological insulators, and the bulk materials of Weyl
semimetals. In particular, the helical spin textures of Weyl
semimetals manifest the spin-momentum locking of Weyl fermions in a
visible manner. The spin-wave fluctuations of the helical order carry
electric charge density; therefore, the spin textures can be
electrically controlled in a simple and predictable manner.

\end{abstract}


\maketitle

Relativistic electrons governed by the Dirac equation had been
thought to be remote from condensed matter physics. The developments
in the last decade, especially the discovery of
graphene\cite{geim2007rise} and topological insulators
(TIs)\cite{qi2010a,hasan2010,qi2011}, however, have established the
ubiquitousness of Dirac fermions in condensed matter. The
topologically protected surface states of topological insulators are
generally massless Dirac fermions, meanwhile, the bulk states of many
topological insulators can be described by massive Dirac equations.
More recently, massless
Dirac\cite{liu2014discovery,neupane2014,Borisenko2014,xu2015observation}
(and
Weyl\cite{wan2011,volovik2003,burkov2011,zyuzin2012a,witczak2012,hosur2012,
aji2011,liu2012,
xu2011,yang2011,wang2012e,halasz2012,jiang2012,delplace2012,meng2012,
garate2012,grushin2012,son2012}) fermions have also been discovered
in bulk materials ( a recent experimentally discovered material is
the TaAs
class\cite{weng2015,Huang2015TaAs,Zhang2015a,xu2015,lv2015,Huang2015,Ghimire,Shekhar,Xu2015NbAs}
).

The interactions among the nominal massless Dirac fermions, if sufficiently strong, can dynamically generate a Dirac mass and fundamentally change the properties of fermions. This phenomenon was first studied in the context of particle physics\cite{nambu1961}.  In this paper we investigate the physical consequences of dynamical mass generation in several classes of topological Dirac metals. We find that the dynamically generated masses manifest themselves as helical spin orders. These types of spin order have attracted considerable interest in other materials and models\cite{nakanishi1980origin,ishikawa1976helical,uchida2006real,
binz2006theory,bernevig2006exact,koralek2009emergence}. As we shall show, their emergence in topological Dirac semimetals is a quite robust phenomenon, which is independent of material details. Physically, the helical spin orders result from the spin-orbit coupling. Their descriptions as Dirac fermions permit a unified treatment.

\begin{figure}
\includegraphics[width=8.0cm, height=1cm]{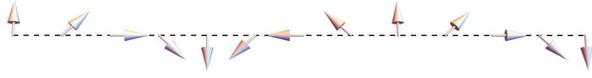}
\caption{ Helical order at the edge of QSH. } \label{QSH}
\end{figure}

We present the results for three classes of materials. The first example is the edge state of quantum spin Hall (QSH) insulators\cite{kane2005b,bernevig2006c,konig2007}. Dynamically generated Dirac mass manifests itself as helical order at the edge [see Fig.\ref{QSH}]. Quantum fluctuations, however, can destroy this order. Proximity to other materials (e.g. proximity to certain superlattice structures) can stabilize the helical order. The second example is the surface states of weak topological insulators, which are closely related to the first example, while avoiding the strong quantum fluctuations because of the higher dimensionality.
The third example is
Weyl semimetal in the magnetic field. Here, the helical order depends on the direction of the magnetic field in a specific manner, which provides a sharp experimental signature for the identification of Weyl semimetals\footnote{A closely related phenomenon called ``chiral magnetic spiral'' has been investigated in the context of quark matters, see Ref.\cite{Schoen2000,Basar2010,Basar2009,kojo2010quarkyonic}.}. The second and the third examples will be our focuses.


\emph{Helical order at QSH edge.}  The edge of QSH accommodates two counterpropagating modes, whose spin
is locked with the propagating direction\cite{wu2006helical,xu2006stability}. The Hamiltonian reads \bea
H(k)=\sum_k v_F(k-k_F)c^\dag_{\uparrow k} c_{\uparrow k} -\sum_k
v_F(k+k_F)c^\dag_{\downarrow k} c_{\downarrow k} \eea where the
chemical potential has been absorbed into the definition of $k_F$.
This form of $H(k)$ is dictated by time reversal symmetry. We can
introduce $c_{R/L}$ by $c_\uparrow (x) = e^{ik_Fx}c_R(x)$ and
$c_\downarrow (x)=e^{-ik_Fx}c_L(x)$, then the Hamiltonian becomes
\bea H(p) =v_F\sum_p p c^\dag_p \sigma_z c_p, \eea where $c_p\equiv
(c_{Rp}, c_{Lp})^T$ is the Fourier transformation of $c_{R/L}(x)$,
and $\sigma_z$ is the spin operator. This Hamiltonian describes a one-dimensional massless Dirac metal (more precisely, a Weyl metal), with $\sigma_z$ playing the role of chirality.

In this paper we focus on the possibility of
dynamical fermion mass generation and symmetry
breaking\cite{fradkin1983,wang2013a,wei2012}. For simplicity let us take the interaction to be short-ranged, namely, $H_I = -g(c^\dag_L c_R) (c_R^\dag c_L)$ (with $g>0$, which means repulsive interaction among electrons with opposite chirality). In the mean field theory, we define $m=g\la c^\dag_L c_R\ra$, and obtain its mean field value as $|m|= m_0\equiv v_F\Lambda \exp(-\frac{2\pi v_F }{g})$, where $\Lambda$ is a momentum cutoff. In this mean field calculation, only $|m|$ can be obtained, while the phase of $m$ is arbitrary. It is convenient to write $m(x)=m_0\exp(i\theta(x))$, neglecting the fluctuation of $|m|$. The ground state without Goldstone modes is $\theta(x)=\theta_0$, where $\theta_0$ is an arbitrary real constant.




The expectation of the $x$-component of spin is given by \bea \la\sigma_x(x)\ra  &=& e^{-iQx} \la c_L^\dag(x)c_R(x) \ra + e^{iQx} \la
c_R^\dag(x)c_L(x)\ra \nn \\
&=&  2\frac{m_0}{g} \cos(Qx-\theta(x)) \eea where $Q\equiv 2k_F$. Similarly, we have \bea
\la\sigma_y (x)\ra &=&   -i e^{-iQx} \la
c_L^\dag(x)c_R(x)\ra +  i e^{iQx} \la
c_R^\dag(x)c_L(x)\ra \nn \\ &=&  -2\frac{m_0}{g} \sin(Qx-\theta(x)) \eea and $\la\sigma_z(x)\ra = \la c^\dag(x)\sigma_z c(x)\ra  =0$.
The helical order is illustrated in Fig.(\ref{QSH}).
So far a physically intuitive mean-field analysis is presented. A more rigorous approach is the bosonization. The helical spin ordering is favored in the presence of sufficiently strong interaction and certain perturbation enabling the umklapp processes (see the online Supplemental Material).



\emph{Helical order at the surface of weak TI.} A natural recipe to avoid strong quantum fluctuations in one-dimensional (1D) is to couple many 1D systems together to form a 2D system. This picture brings us to the study of this section.

\begin{figure}
\centering
\includegraphics[width=9cm, height=6cm]{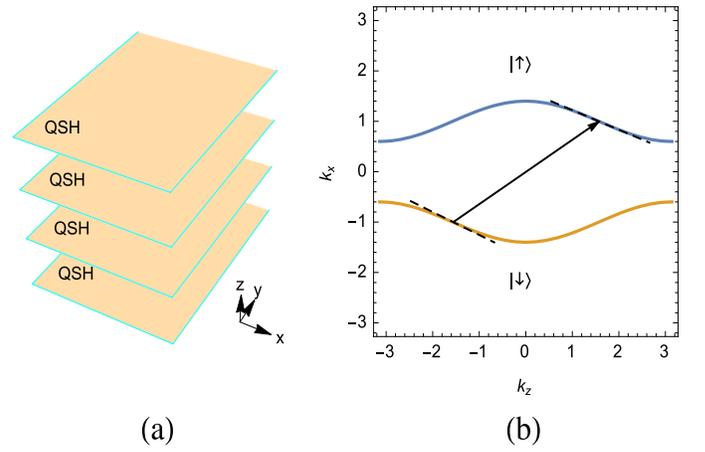}
\caption{ (a) Weak TI as layered QSH. (b) Fermi surface nesting at the surface (in the $xz$ plane) of weak TI. } \label{nesting}
\end{figure}

Weak TIs are characterized by the so-called weak topological indices\cite{fu2007b,Yan2012Prediction}. The simplest models of weak topological insulators consist of layered QSH  [see Fig.\ref{nesting}]. Therefore, the surface states can be obtained by coupling the QSH edge states. Suppose that the surface of the weak TI coincides with the $xz$-plane. To simplify the problem, we include hopping terms among only adjacent layers. The low energy Hamiltonian for the surface reads
\bea H(\bk)&=&\sum_\bk v_F(k_x-k_F)c^\dag_{\uparrow \bk} c_{\uparrow \bk} -\sum_\bk v_F(k_x+k_F)c^\dag_{\downarrow \bk} c_{\downarrow \bk} \nn \\ && -2t_\parallel\sum_\bk \cos k_z (c^\dag_{\uparrow \bk} c_{\uparrow \bk} + c^\dag_{\downarrow \bk} c_{\downarrow \bk}), \eea where $\bk\equiv (k_x,k_z)$, and $t_\parallel$ is the interlayer hopping.

\begin{figure}
\includegraphics[width=6.0cm, height=6cm]{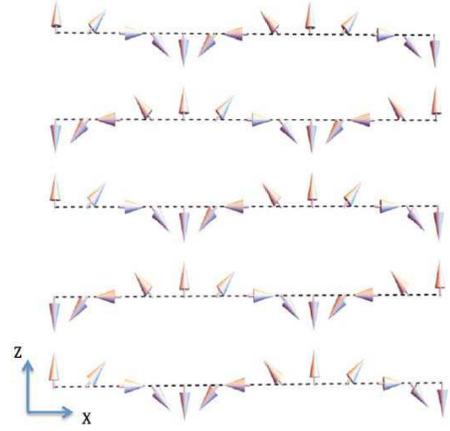}
\caption{ Helical order at the surface of weak TIs. This spin order is helical in the $x$ direction and staggered in the $z$ direction. } \label{Weak_TI}
\end{figure}

A notable feature of this Hamiltonian is as follows. The energy of electrons with spin up and spin down is $E_\uparrow (\bk)=v_F(k_x-k_F)-2t_\parallel \cos k_z$ and $E_\downarrow (\bk)= -v_F(k_x+k_F)-2t_\parallel \cos k_z$, respectively, therefore, we have \bea E_\uparrow (\bk+\bQ)=v_F(k_x+k_F) + 2t_\parallel \cos k_z =-E_\downarrow (\bk) \eea where $\bQ\equiv (2k_F, \pi)$. Therefore, there is perfect Fermi surface nesting at wavevector $\bQ$ (This salient feature is absent in the surface states of the strong topological insulator\cite{jiang2011spin}). An infinitesimal interaction $H_I = -g(c^\dag_L c_R) (c_R^\dag c_L)$ can generate a gap, analogous to the case of the QSH edge. The order parameter of this symmetry breaking is $m(\bx)=g\la c^\dag_L (\bx) c_R(\bx)\ra \equiv m_0\exp(i\theta(\bx))$. The spin densities are given by \bea \la\sigma_x(\bx)\ra &=& e^{-i\bQ\cdot\bx} \la c_L^\dag(\bx)c_R(\bx) \ra + e^{i\bQ\cdot\bx} \la
c_R^\dag(\bx)c_L(\bx)\ra \nn \\
&=&  2\frac{ m_0}{g} \cos(\bQ\cdot\bx-\theta(\bx)) \eea
where $Q\equiv (2k_F,\pi)$, and \bea
\la\sigma_y (\bx)\ra &=&   -i e^{-i\bQ\cdot\bx} \la
c_L^\dag(\bx)c_R(\bx)\ra +  i e^{i\bQ\cdot\bx} \la
c_R^\dag(\bx)c_L(\bx)\ra \nn \\ &=&  -2\frac{m_0}{g} \sin(\bQ\cdot\bx-\theta(\bx)) \eea The spin texture at the weak TI surface is illustrated in Fig.\ref{Weak_TI} (We have taken $\theta(\bx)$ to be a constant). The spin order is helical in the $x$ direction, while staggered in the $z$ direction. Finally, we remark that in real materials the perfect nesting is replaced by approximate nesting. To be more conclusive, we have studied a realistic lattice model in the random phase approximation (RPA), and found that the approximate nesting favors helical spin order even if the interaction is quite weak(see Supplemental Material), therefore, we expect it to occur in real materials of the weak topological insulator.


\emph{Helical spin order in Weyl semimetals.} The dynamical mass
generation induced by interaction and the resultant charge density
wave state has been studied before\cite{wang2013a,wei2012,zyuzin2012,maciejko2014,zhang2015,buividovich2014}. With
an external magnetic field, the Fermi surface instability becomes
infinitesimal, i.e. an infinitesimal interaction can open up a gap at
the Fermi surface\cite{yang2011,roy2014}. The charge density wave pattern is, however, too crude to identify the unique feature of Weyl semimetals, namely, the spin-momentum locking described by the Weyl equation. Here we show the existence of helical spin orders, which provides a finer signature of the Weyl-type spectrum.

For simplicity let us consider a single pair of Weyl points located
at $\bK_1$ and $\bK_2$ respectively. We define the shorthand notation
$\bQ=\bK_1-\bK_2$. The low-energy Weyl Hamiltonian reads $h_s(\bp) =
v_F \sum_{i,j=x,y,z}p_i e_{ij}^s\sigma_j $, in which $s=1,2$,
$p_i\equiv k_i-K_{s,i}$, and $e^s$ is a $3\times 3$ matrix. For each
$i=x,y,z$ we can define a vector
$\be^s_i=(e^s_{ix},e^s_{iy},e^s_{iz})$; thus $h_s$ becomes more
compact: \bea h_s(\bp) = v_F \sum_{i=x,y,z} p_i
(\be_i^s\cdot\bsigma), \label{weyl-H} \eea where $p_i$ is small
compared to $|\bQ|$. Hereafter we shall focus on the cases with
$\be^s_i\cdot\be^s_j=\delta_{ij}$, for which compact analytical
treatment is possible.

Now we add a magnetic field $\bB=B\hat{z}$ (We can always rotate the
coordinate system such that $\bB$ points in the $\hat{z}$ direction)\footnote{ We
focus on the orbital effects of magnetic field, omitting the Zeeman
energy at this stage.}.
The energy eigenvalues for nonzero Landau levels are \bea E_l( p ) =
\pm v_F \sqrt{p_z^2+2eBl} \,\,\,\, (l=1,2,3,\dots), \eea which are all
gapped. To obtain the zeroth Landau level, first we can solve the 2D
problem in the $xy$-plane by letting $p_z=0$. In the Landau gauge the 2D
Hamiltonian is \bea h_{2D}(q_x, q_y) = v_F [(p_x +eBy)
\be_x^s\cdot\bsigma + p_y\be_y^s\cdot\bsigma)]. \eea We can find that
the zeroth Landau level wavefunction $ \psi_{p_x}(x,y)
=\frac{1}{\sqrt{2\pi eB}}\exp[-\frac{1}{2eB}(eBy+p_x)^2]\exp(ip_x
x)|\be^s_x\times\be^s_y\ra$, where we have introduced the notation
$|{\bf n}\ra$ to denote the two-component spinor satisfying $\la {\bf
n}|\sigma_i|{\bf n}\ra = n_i$ for an vector ${\bf n}$. Adding the
$p_z$ term is now straightforward because $|\be^s_x\times\be^s_y\ra$
is an eigenvector of $p_z\be^s_z\cdot\bsigma$ (This is the
simplification of taking $\be^s_i\cdot\be^s_j =\delta_{ij}$). The
single-particle wavefunciton is \bea \psi^s_{p_x,p_z}(\bx)
=\frac{1}{\sqrt{2\pi eB}}\exp[-\frac{1}{2eB}(eBy+p_x)^2]\nn \\
\times\exp(ip_x x + ip_z z)|\chi_s\be^s_z\ra, \eea in which we have
introduced the chirality  \bea \chi_s =
(\be_x^s\times\be_y^s)\cdot\be_z^s =\pm 1, \eea and the
single-particle energy is \bea E^s_{p_x,p_z}  = \chi_s v_F p_z,
\label{dispersion} \eea According to the above wavefunction
structure, the fermion operators can be expanded as \bea
c(\bx)=\sum_s e^{i\bK_s\cdot\bx}|\chi_s\be^s_z\ra c_s(\bx) +\dots,
\eea where $c_s(\bx)$ are low-energy fermion operators analogous to $c_{R/L}(x)$ in the QSH section, and ``$\dots$'' denotes high energy modes. Suppose that $\chi_1=-\chi_2=1$, then the index
identification $1(2)\leftrightarrow R(L)$ is valid, and the analysis
of dynamical symmetry breaking in the QSH section applies, namely, a
four-fermion interaction induces a chiral condensation $\la
c^\dag_L(\bx)c_R(\bx)\ra=\frac{m(\bx)}{g}\equiv
\frac{m_0}{g}\exp(i\theta(\bx))$. The $x$-component of spin density
becomes \bea \la\sigma_x(\bx)\ra &=& \la c^\dag(\bx)\sigma_x
c(\bx)\ra \nn
\\ &=& e^{-i\bQ\cdot\bx}\frac{m(\bx)}{g} \la\chi_2
\be^2_z|\sigma_x|\chi_1\be^1_z\ra + H.c. \eea If we write $|\chi_1
\be^1_z\ra =|\be^1_z\ra= [\cos(\phi_1/2),
\sin(\phi_1/2)e^{i\varphi_1}]^T$ and $|\chi_2 \be^2_z\ra =
|-\be^2_z\ra= [\sin(\phi_2/2), -\cos(\phi_2/2)e^{i\varphi_2}]^T$,
then we have $\la\chi_2 \be^2_z|\sigma_x|\chi_1\be^1_z\ra =
\sin(\phi_1/2)\sin(\phi_2/2)e^{i\varphi_1}-\cos(\phi_1/2)\cos(\phi_2/2)e^{-i\varphi_2}.
$ Similarly, we have \bea \la\sigma_y(\bx)\ra &=& \la
c^\dag(\bx)\sigma_y c(\bx)\ra \nn
\\ &=& e^{-i\bQ\cdot\bx}\frac{m(\bx)}{g} \la\chi_2
\be^2_z|\sigma_y|\chi_1\be^1_z\ra + H.c. ,\eea with $\la\chi_2
\be^2_z|\sigma_y|\chi_1\be^1_z\ra=
-i[\cos(\phi_1/2)\cos(\phi_2/2)e^{-i\varphi_2}+\sin(\phi_1/2)\sin(\phi_2/2)e^{i\varphi_1}]$,
and \bea \la\sigma_z(\bx)\ra &=& \la c^\dag(\bx)\sigma_z c(\bx)\ra
\nn
\\ &=& e^{-i\bQ\cdot\bx}\frac{m(\bx)}{g} \la\chi_2
\be^2_z|\sigma_z|\chi_1\be^1_z\ra + H.c. ,\eea with $\la\chi_2
\be^2_z|\sigma_z|\chi_1\be^1_z\ra=
\cos(\phi_1/2)\sin(\phi_2/2) +\sin(\phi_1/2)\cos(\phi_2/2)e^{i(\varphi_1-\varphi_2)}$. Finally, the charge density is \bea \la\sigma_0(\bx)\ra &\equiv& \la c^\dag(\bx)  c(\bx)\ra
\nn
\\ &=& e^{-i\bQ\cdot\bx}\frac{m(\bx)}{g} \la\chi_2
\be^2_z| \chi_1\be^1_z\ra + H.c. ,\eea with $\la\chi_2
\be^2_z |\chi_1\be^1_z\ra=
\cos(\phi_1/2)\sin(\phi_2/2) -\sin(\phi_1/2)\cos(\phi_2/2)e^{i(\varphi_1-\varphi_2)}$.

\begin{figure}
\centering
\includegraphics[width=8.5cm, height=6.5cm]{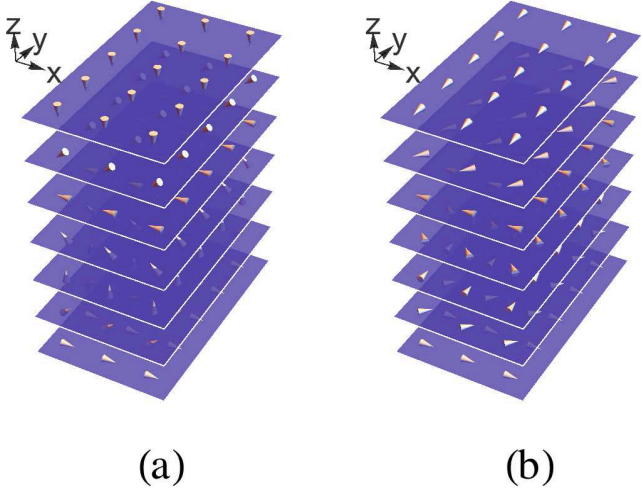}
\caption{ Helical spin orders in Weyl semimetals (with dynamical mass
generation) for (a) Hamiltonian in Eq.(\ref{H-1}) and (b) Hamiltonian
in Eq.(\ref{H-2}). Here $\bQ$ is taken to be in the $\hat{z}$
direction. } \label{Weyl}
\end{figure}

Studying some special cases helps us to understand these results. For
instance, we consider \bea
h_1(\bp)&=&v_F(p_x\sigma_x-p_y\sigma_z+p_z\sigma_y) , \nn \\
h_2(\bp)&=& v_F(p_x\sigma_x+p_y\sigma_z+p_z\sigma_y). \label{H-1}
\eea  It is readily seen that $\be^1_z=\be^2_z= (0, 1, 0)$, and the previous general
results tell us that \bea \la\sigma_x(\bx)\ra &=&
\frac{2m_0}{g}\sin(\bQ\cdot\bx-\theta(\bx)), \nn \\
\la\sigma_y(\bx)\ra &=& 0, \nn \\ \la\sigma_z(\bx)\ra &=&
\frac{2m_0}{g}\cos(\bQ\cdot\bx-\theta(\bx)). \eea This case is shown
in Fig.(\ref{Weyl}a). If we take a different Weyl Hamiltonian \bea
h_1(\bp)&=&v_F(p_x\sigma_x+p_y\sigma_y+p_z\sigma_z), \nn \\
h_2(\bp)&=& v_F(p_x\sigma_x-p_y\sigma_y+p_z\sigma_z), \label{H-2}
\eea the a simple calculation yields  \bea \la\sigma_x(\bx)\ra &=& -
\frac{2m_0}{g}\cos(\bQ\cdot\bx-\theta(\bx)), \nn \\
\la\sigma_y(\bx)\ra &=& -\frac{2m_0}{g}\sin(\bQ\cdot\bx-\theta(\bx)),
\nn \\ \la\sigma_z(\bx)\ra &=& 0. \eea This case is shown in
Fig.(\ref{Weyl}b). The charge density $\la\sigma_0(\bx)\ra=0$ for
these two cases.  Eq.(\ref{H-1}) and Eq.(\ref{H-2}) qualitatively
resemble the Weyl semimetal materials and models. For instance, the
simple model\cite{yang2011} $h(\bk)=2t_x\sin k_x\sigma_x + [2t_y(\cos
k_y-\cos k_0)+m(2-\cos k_x -\cos k_z)]\sigma_y  +2t_z\sin
k_z\sigma_z$ has a pair of Weyl nodes at $( 0, \pm k_0, 0)$, with
$h_{s=1,2}= v_x\sigma_x p_x   \pm v_y \sigma_y p_y + v_z \sigma_z
p_z$ as its low-energy approximation, which is the same as
Eq.(\ref{H-2}) except for the possible velocity anisotropy. A more
quantitative study of these materials shall be presented elsewhere.

By changing the direction of the magnetic field, $|\be^s_z\ra$ is changed
accordingly, and the helical spin texture changes in a prescribed
way. We also remark that if the Pauli
matrices in Eq.(\ref{weyl-H}) are not associated with spin but some
other degrees of freedom, the helical order is a straightforward
generalization of the above results.

The spin helix predicted here can be observed by the spin-resolved
scanning tunneling microscope (STM). It is unclear whether the
magnitude of electron-electron interaction (and the sample quality)
of the recently discovered Weyl semimetals favors the generation of the
spin helix, but we are probably justified in being optimistic about its
possible realization, considering the ongoing rapid progress in this
field.


\emph{Electric manipulation of spin texture.} So far we have not investigated the effects of
fluctuations of $\theta$.
Such phase fluctuations are termed
``axions''\cite{peccei1977,wilczek1978problem,weinberg1978new}, and
have also been studied in the context of topological insulators and superconductors\cite{qi2008,li2010,wang2011b,ryu2012,wang2012d,qi2013axion}. In our present study, the axions are much more visible because of their simple geometrical meaning: they are the phase angle of spin polarization (rotated from $\bQ\cdot\bx$).

Now we shall show that spin textures carry electric charge; moreover, the charge density depends on the spin texture in a precise manner. For concreteness, let us take the QSH edge as an example. In the presence of $\theta(\bx,t)$, the spin polarization is pointing to angle $\bQ\cdot\bx-\theta(\bx,t)$. Let us consider the simplest case, $\theta=\theta_0-A\cos(qx)$, where $A<<1$ is the amplitude of the spin modulation on the background of the helical order. For such a slow modulation of the phase $\theta$ of the Dirac mass, the Goldstone-Wilczek formula\cite{goldstone1981} relates the gradient of $\theta$ to the charge density: \bea \rho(x)=\frac{1}{2\pi}\partial_x\theta =\frac{A}{2\pi}q\sin(qx). \eea On the other hand, if the charge density is given as $\rho(x)=\rho_0\cos(qx)$, then we have \bea \theta(x)=\frac{2\pi\rho_0}{q}\sin(qx). \eea Therefore, gating the system periodically, such that the charge density modulates periodically, can control the spin modulation in a predicable manner.

We can also consider gating the system to induce a constant charge density $\rho_0(\bx)=C$. According to the Goldstone-Wilczek formula, we have $\theta(\bx)=2\pi Cx$ for this case. Now the phase $Qx-\theta(x)=(Q-2\pi C)x$, namely, the wavevector of helical spin order becomes $Q-2\pi C$. This is consistent with the relation $Q=2k_F$: a constant charge density amounts to shifting $k_F$ in the underlying Fermi surface ``before'' dynamical mass generation.

Finally, we remark that taking $k_F=0$ brings us back to the result
of Ref.\cite{qi2008b}; namely, a magnetic domain wall between $+x$
and $-x$ magnetization generates a fractional charge $\pm e/2$. For a
general $k_F\neq 0$, the $\pm x$ magnetization is replaced by a spin
helix with a wavevector $Q=2k_F$, i.e. both sides of the domain wall
are spin helices, with a phase difference $\pi$.


\emph{Conclusions.}  A most prominent feature of topological Dirac
and Weyl semimetals is the spin-momentum locking, which is a
dramatic consequence of spin-orbit coupling. We have shown that this
spin-momentum locking can be ``frozen''  as helical spin ordering in
the presence of dynamical instability  (``helical solids'' from
helical liquids). These real-space (not the reciprocal-space) helical spin textures should be visible in the spin-resolved STM. Apart
from its intrinsic interest, it has potential applications due to the
electric tunability of helical spin texture.

\emph{Acknowledgements.} Z.W. is supported by NSFC under Grant No. 11304175 and the Tsinghua
University Initiative Scientific Research Program. S.C.Z. is supported
by  the NSF under Grant numbers
DMR-1305677 and the US Department of Energy, Office of Basic Energy Sciences under
Contract No. DE-AC02-76SF00515.

\bibliography{dirac}

\end{document}